\documentclass[english]{sobraep}
\usepackage{blindtext}
\usepackage{hyperref}

\title{Eff the ineffable: on the uncommunicability of a conceptually simple contribution to HCI methodology}

\author{Helen Oliver$^{1}$,$^{2}$, Richard Mortier$^{2}$, Jon Crowcroft$^{1}$,$^{2}$\\
	\normalsize $^{1}$The Alan Turing Institute, London -- UK\\
	\normalsize $^{2}$University of Cambridge, Cambridge -- UK\\
	\normalsize e-mail: helen dot oliver at bravo charlie sierra dot oscar romeo golf dot uniform kilo, richard.mortier@cl.cam.ac.uk, jac22@cam.ac.uk
}

\begin{document}

\maketitle

\begin{abstract}
Given a simple concept that has the potential for a methodological contribution to the field of HCI, the constraints of submission conventions within the field may make it impossible to communicate the concept in a manner that is intelligible to the reader.
\end{abstract}


\textit{Authors' note} \footnote{This article was written especially for AltChi2022, a track for papers that would be inherently unsuitable for the main CHI conference.
\linebreak Our submission received five peer reviews, all top-rated and all strongly recommending acceptance; but the paper sadly did not survive the jury's deliberations.  \linebreak We do not consider this paper suitable for submission to any other venue. \linebreak However, the enthusiastic response from the peer reviewers indicates that it has some value to offer to the research community, so we are adding it to arXiv.org as a preprint/e-print. \linebreak This version of the article has been typeset with a non-ACM template (sobraep.cls). The only other change from the submitted version is the inclusion of Richard Mortier as second author, who had been omitted because of a technical issue with HotCRP.}

%~~~~~~~~~~~~~~~~~~~~~~~~~~~~~~~~~~~~~~~~
%Sections
%~~~~~~~~~~~~~~~~~~~~~~~~~~~~~~~~~~~~~~~~

%Introduction

\section{INTRODUCTION}

I am going to introduce this paper by stating what actually inspired my research, and the reasoning behind my choices. In order to do this, I will have to write in a style that goes against the conventions of a scholarly article on Human-Computer Interaction. I have to do this because, although the concept is simple, the editorial constraints of the submission process have convoluted my efforts to communicate it.

In this section I will describe the inciting incident that inspired my research, the decisions I made in order to act on that inspiration, and the reasoning behind those decisions. 

In 2014, I went to a meeting of the Wearables Special Interest Group at Cambridge Wireless \footnote{\textit{Cambridge Wireless - Connected Devices Group,} 9 October 2014: \textbf{Wearables: The Internet of Us.} \href{https://www.cambridgewireless.co.uk/events/46845-connected-devices-sig-event-wearables-the-in/}\linebreak{https://www.cambridgewireless.co.uk/events/46845-connected-devices-sig-event-wearables-the-in/}}. 

At that meeting, Nick Hunn, an industry expert, spoke about the blandness of consumer wearables \cite{hunn2014the}: ``The average street [fashion] market has more variety than the entire wearable tech market.'' Hunn's white paper \cite{hunn2015the} points out the dominance of technology push, and the lack of emotional engagement. 

A show of hands, in a roomful of wearable technology enthusiasts, revealed that the number of those present who actually \textit{owned} any wearable technology could be counted on the fingers of one hand, and fewer than that were wearing theirs at the time.

This made an impression on me. My raison d’\^etre is finding new things to wear, and yet I had not (and still have not) found a single item of wearable technology worth purchasing. As a computer scientist, I would have expected to be an early adopter.

At that time, in 2014 and 2015, I went to a number of meetings, at which the speakers and attendees rallied round the need for a more user-centred, less predictable \cite{mcgookin2014wearables} approach to wearables design. At a one such event, Maciocci, another industry expert, described a general failure of mainstream commercial wearables to offer much of interest to users \cite{maciocci2013me}. 

At that Cambridge Wireless meeting, one of the venue's administrative staff told me, in the ladies' room, that she had enjoyed that day's meeting the most out of all the events she had been involved in, that she had found it so inspiring.

I asked her what wearable device she would choose, if she could have anything she wanted. She searched the air with her eyes, looking for the will-o'-the-wisp of the idea she knew was there but could not catch or see.

Of course she could not answer, because just asking ``SO WHAT WEARABLE TECH DO YOU WANT ME TO MAKE'' is not a productive way of posing the question. How could I ask that question in a way that would get a meaningful answer?

\section{BACKGROUND}

At the expert forums I described in the previous section, user-centred design was proposed as the solution to the apparent problem of boring wearable technology. User-centred design means the designer comes up with an idea or prototype, then takes it to users to get their reactions and feedback to inform further development.

But if the problem is that my ideas for wearables are boring \cite{maciocci2013me}, thinking up another idea to bring to the users is not necessarily the solution. The next idea might also be boring.

Rather than user-centred design, this sounds like a job for participatory design, overlapping with co-design, both of which give the users far more opportunity for input. 

But even participatory design usually starts with a design exemplar or preselected type of device, which is given to participants in search of creative input. One example of a participatory study that gives the participants a substantial amount of creative control is Pateman et al. \cite{pateman2018the}, who asked participants to design their own low-fidelity prototypes for an activity monitor. In that study, the participants were already users of activity monitors. However, most such studies, be they user-centred or participatory, are based on technology push. This means the devices being studied are not necessarily ones the participants would otherwise want to use. Even some of the participants in \cite{pateman2018the}, wearing their own designs for a type of device they already knew they liked, found that the prototypes they had made did not work for them in reality. The proof of the wearable is in the wearing. 

So, firstly, if wearables are boring, involving users or participants in the process of designing variations on an existing boring theme is not the solution. And, secondly, the proof of a wearable is in the wearing - it is not enough to get creative input from participants unless they can test out (some version of) their ideas in reality.

\section{A SIMPLE PROPOSAL}

If wearables are boring, let participants be the ones to provide the idea. We could then build the closest real-world implementation of that idea we could manage, and give it to them to wear in-the-wild. 

But, as I already established in section 1: I cannot just ask users point-blank, because asking point-blank does nothing to help them to imagine wearables that might not exist yet.

In 2012, Andersen and Wilde \cite{andersen2012circles} did a paper prototyping workshop that used embodied techniques to help participants to ``sneak up on themselves'', as they put it: they listed a set of emotion-words, got the participants to pick one, asked them where on their body that emotion was located, and went from there. 

That study showed evidence of users' potential for creativity, a potential which is often dismissed by designers \cite{mueller2005hug} \cite{vanerp2018think}. However, those paper prototypes were not meant to become real functioning devices. And methods involving ``sneaking up on'' oneself are more indirect than I wanted to be (even though the actual methods as described were more direct than the phrase implies). 

I wanted to know the DIRECT answer to the question ``what wearable device do you want?''. I wanted to find the most direct way of asking that I possibly could, but it had to support the participants' imagination.

I decided to get the participants to tell me a story. It is hard to tell a story without emotion, and the more emotional engagement, the less boredom (I hoped). A story of a blue-sky wearable is also a direct answer, and more than that, it is a richly detailed answer which tells the listener a great deal about the storyteller.

We live in a society, so I specified that the participants must not only describe a wearable device but describe the character who wears it and the world the character lives in. Wearables are a form of ubiquitous computing, after all. Dourish and Bell \cite{dourish2014resistance} wrote a whole paper about science fiction's worldbuilding as contextualization of the technologies they describe. For true insight into the participants as potential users of the technologies they desired, we required not just stories, but story\textit{worlds}. In this way, the participants would be able to contribute to the conversation, not just ideas for wearable technologies in isolation, but insights about the kinds of sociotechnical futures they wanted to work towards.

Storytelling is hard, so the least I should do was give them a prompt. Time is short and attention is shorter in this economy, so I decided I would ask them to keep it to five minutes, and give them some pointers about what to cover in that time. Johnson \cite{johnson2011science} has helpfully enumerated ``Five Steps'' to creating a science fiction prototype - those five steps, it seemed to me, would provide a most practical recipe \cite{kymalainen2015evaluating} containing every ingredient that goes into a ubicomp story; or a ``Design Fiction'' \cite{bleecker2009design} \cite{sterling2009design} as they call it (DF for short).

So that is the story of how I arrived at the novel implementation of Participatory Design Fiction (PDFi) that I used in my study of obstacles to wearable computing \cite{oliver2021obstacles}. It is a simple story: wearables are boring, so let's ask people what wearables they would want if they could have anything; then build them something real to wear, and ask them for feedback about the experience, and what kind of future world they imagine living in while wearing it. Everything else is a mere elaboration of that simple central idea.

\section{NOVELTY AND CONTRIBUTION}
This was the first ever study to use PDFi to elicit concepts:

\begin{itemize}
\item{from independent adult participants (so they would have agency to participate as they wished, without being overridden by parents \cite{edwards2011exploring}, carers, bosses or other authority figures \cite{southern2014imaginative}),}
\item{of real wearable technology probes \cite{hutchinson2003technology} (devices with minimum functionality to test the experience of wearing the device in reality, and as a perceptual bridge \cite{auger2012why} between the real world and the storyworld),}
\item{for everyday connected wearables \cite{oliver2019design} (i.e wearables for everyday situations, that send or receive data, and are worn solely out of choice and not necessity - this meant no medical, safety-critical, or occupational devices; and no devices with electronic components but without connectivity),}
\item{with an open-ended brief (to give editorial control and imaginative space to participants),}
\item{for in-the-wild study (to validate the ideas and develop the devices),}
\item{and for co-imagining sociotechnical futures for everyday connected wearables (an attempt at contributing to the democratization of this emerging technology) \cite{oliver2021obstacles}.}
\end{itemize}

At this point, you are probably asking one or more of the following three questions:

\begin{enumerate}
    \item What is the methodology?
    \item What is novel about this work?
    \item What is the contribution?
\end{enumerate}

\subsection{What is the methodology?}
This is not a methodology paper, and I will not describe my methodology at all here. Space does not permit me to go into sufficient detail within the current article, so the reader is referred to \cite{oliver2021obstacles}, Chapters 1--3, Appendices H, I and K. The purpose of this article is to describe the systemic barriers I encountered to my attempts at communicating the simple reasoning underlying the methodology.

\textbf{Not good enough. Every paper has to stand on its own.} One reason I am not going to describe my methodology in this paper is because my every attempt to do so until now has been frustrated by the editorial process. Part of the problem is anonymity, which does not apply here.

Anonymity policies make it very difficult to build on previous work. At the end of my position paper \cite{oliver2019design}, I declared that my next article would be an in-depth thematic analysis of the PDFis. That would not, itself, be a methodology paper. But there is still a requirement to describe the methodology in sufficient detail that anyone could repeat it based on the content of the given paper alone. Without that context, reviewers are likely to find the paper incomprehensible.

\subsubsection{Anonymity and the disorientation of Review Panel Ping-Pong}
When I submitted the thematic analysis paper, one of the reviewers rejected it as a mere repetition of material I had covered before. On the one hand, I totally disagreed with the accusation that I had repeated myself; I had touched on some of the same material on one occasion when speaking, but speaking is not publication, nor had I gone into nearly as much depth. On the other hand, I did not want to lay myself open to any further accusations of repeating myself. So I rewrote the paper from the ground up, with a completely different framing, and playing ``the floor is lava'' \footnote{\begin{flushleft} \textit{Wikipedia:} \textbf{The Floor Is Lava} \href{https://en.wikipedia.org/wiki/The_floor_is_lava}{\linebreak https://en.wikipedia.org/wiki/The\_floor\_is\_lava} \end{flushleft}} with every word I had ever committed to print until then. 

To this brand new paper, one reviewer objected that it should so obviously have been an in-depth thematic analysis, and why had I not done that instead? Plus, it was strangely lacking in methodological context, and needed more explanation. After several days of formatting it to fit the rebuttal form, my offer to incorporate the methodological background and thematic analysis was of course rejected, because of course that would make the paper too different. 

The problem of ping-ponging between opposite demands with each successive set of reviewers is a recognized phenomenon, and one which already has a proposed solution: keeping the same set of reviewers across multiple venues. So I will not recount my experiences of games three, four and five of Review Panel Ping-Pong. Nor do the constraints of the situation diminish my own responsibility for clear and communicative writing; in particular, I did not agree with the accusation of repeating myself, so I should not have been reactive to it and arguably set myself up to fail at the outset. 

But Review Panel Ping-Pong is disorienting and destabilizing for the author playing the role of the ball, and it becomes more and more difficult to know whether one is conveying enough of the right information with each successive attempt. 

\subsubsection{Anonymity, detail, and the problem of Matrioshka Unpacking}
An example of one occasion when disorientation, resulting from an overload of contradictory feedback, led me to communicate less clearly than I realized, was the time a reviewer complained that I had underspecified my process because I had described conducting four pilot workshops. Why did I not stipulate whether, in order to repeat the method, someone else would also have to do four pilot studies? The word ``pilot'' was not conveying the connotations I imagined it was conveying, of a trial of that part of the method itself. Having put the workshop format to the trial, it should not be necessary for subsequent adopters to do so again - and I did not realize I needed to say so explicitly. 

As an author one would hope to keep descriptions of previously published work brief, but because self-citations (if any) in anonymized papers are required to be so discreet as to pass unnoticed, it is difficult to build on previous work. On one occasion, a reviewer did call me out for deanonymizing myself even though I had followed the prescribed format to the letter; they suggested briefly contextualizing the previously published work with a diagram. I did so, adding a half page of description explaining the diagram. This was not enough detail to satisfy the other reviewers. I added some more detail. It was still not enough detail. More reviewers demanded more detail, and even more reviewers demanded even more detail, and a world of reviewers demanded a world of detail, and a universe of reviewers demanded a universe of detail.

Eventually, I ended up rewriting \textit{all} the information from the previous paper in such a way as to expand it into the new paper in its entirety, like unpacking a matrioshka doll which itself is enclosed within a new, larger doll containing every detail demanded by every reviewer, but \textit{without} actually \textit{repeating,} in letter or spirit, anything previously published. 

That is not as easy as it sounds. 

And once I had accomplished that feat, there was not enough space to fit the material for the paper I was actually trying to write.

It is my responsibility to communicate my research clearly and rigorously to the reader. Review Panel Ping-Pong and Matrioshka Unpacking, two phenomena which are unintended consequences of the anonymous review process, make it more difficult than it should be, both to communicate, and to expand on an essentially simple idea.

\subsection{What is novel about this work?}
An important limit on the scope, and one that I cannot emphasize enough (no, really, no matter how much I try to emphasize this point it never seems to come across) is that this was \textit{not} a participatory design study that \textit{happened} to use everyday connected wearables as a motivating example. It was \textit{a study of everyday connected wearables using PDFi \cite{oliver2021obstacles}.} The PDFi was used for inspiration, consequential speculation \cite{elsden2017on}, co-imagining, and reflection. 

Since I started this study, in 2016, a number of articles on PDFi have been published, but the concept was already established at that time. In 2015, Prost et al. \cite{prost2015from} had done a study about sustainable energy, in which participants were given a very specific brief to imagine: the energy consumption habits of a family named Gruber in 2039.

Tsekleves et al. 2017 and 2019 \cite{tsekleves2017codesigning} \cite{tsekleves2019rethinking}, N\"{a}gele et al. 2018 \cite{nagele2018pdfi}, Candello et al. 2019 \cite{candello2019teaching} and Desjardins et al. 2019 \cite{desjardins2019bespoke}, among others, have also done studies of PDFi which gave some degree of editorial control to the participants. 

Tsekleves et al. \cite{tsekleves2017codesigning} \cite{tsekleves2019rethinking} were not about wearables, but the project (called \textit{ProtoPolicy}) did even produce a design fiction in the form of a wearable artefact \footnote{It was a smartwatch for auto-euthanasia, devised by the researchers and consultants, in a study which involved elderly participants to demonstrate the potential of PDFi in public policy. This study is often held up to challenge the novelty of my own study, but it differs in a number of important ways, including that auto-euthanasia is a safety-critical and medical, rather than an everyday, activity.}. In my opinion, however, the study, as described, reads as being more user-centred than participatory. 

Candello et al. \cite{candello2019teaching} provided the beginning of a story - about robot floor guides in museums - and asked participants to complete it. 

Desjardins et al. \cite{desjardins2019bespoke} did a study which involved, among other things, booklets with fictional concepts that included illustrations of participants' own homes, onto which they could draw new objects or layouts, and/or write in their ideas and suggestions. 

The closest PDFi study to my own, however, was N\"{a}gele et al. 2018 \cite{nagele2018pdfi}, who gave an open-ended (science fiction) storytelling brief to patients with chronic urinary tract infections (UTIs). Like my own study, they gave participants a great deal of editorial control at the outset. They then worked iteratively with participants to refine the stories and make imaginary future medical devices. The study differed from my own in several important respects: the imaginary future medical devices, which the researchers fabricated by 3D printing, were by definition not for real-world use; were not wearables; and also, being medical devices, were both safety-critical, and to be used out of necessity rather than free choice. What the two studies have in common is the goal of facilitating the participants to tell their own stories in their own words. Both studies were consequential for participants - in my case, because of the possibility of wearing a real version of one of the fictional wearables; in N\"{a}gele et al's case, because the findings were delivered to a medical device company in order to inform future development of treatments. 

A study which did not involve PDFi, but which is closer still to my own, is Jones et al. \cite{jones2016beyond} who used a ``magic thing'' probe to elicit concepts from emergent users in Bangalore, Nairobi and Cape Town. Nominally, the prompt took the form of a wristband, but because of the open-ended brief and the future orientation of the discussions, participants' ideas - mostly for wearables - were not limited by the wristband form factor.

\textbf{So it's not novel, then. You just said so yourself: PDFi has been used before \cite{prost2015from} \cite{tsekleves2017codesigning} \cite{tsekleves2019rethinking} \cite{nagele2018pdfi} \cite{candello2019teaching} \cite{desjardins2019bespoke}, \textit{and} it's not the first time someone has given independent adult participants an open-ended brief to imagine concepts for wearables \cite{andersen2012circles} \cite{jones2016beyond}. In fact, every method you say you used - creative toolkits \cite{sanders2012convivial}, technology probes \cite{hutchinson2003technology}, you name it - has been done before!} Nobody has done a study like the one described in section 4, in this way, and for these reasons. Jones et al. \cite{jones2016beyond} gave an open-ended brief to independent adult users, and generated actionable insights, but with no intent to make real devices in the immediate term. The level of editorial control given to participants, the open-ended brief, AND the simultaneous current and future orientation of the use of PDFi makes our work unique in the domain of everyday connected wearables for a cohort of independent adult participants recruited from the general public \cite[section 3.1]{oliver2021obstacles}. 

\subsection{What is the contribution?}
PDFi offers potential users of everyday connected wearables to provide a rich description, in their own words, about what they truly want from the technology, how they imagine its enhancing their everyday lives, and what kinds of futures they want to create with their actions in the present.

In Bray and Harrington's \cite{bray2021speculative} work on Afrofuturism, in which they used card methods for participatory speculation in Black and brown communities, they explained: ``Participatory design is positioned as an approach to address community challenges and reimagine potential futures by decentralizing systems of power and centering the marginalized, moving away from designers as experts and framing design as being community-driven. Methods of speculative participatory design are useful in imagining futures among marginalized groups while negotiating existing societal constructs.''

Although I did not seek out marginalized populations, a number of the PDFis were inspired by their storytellers' experiences of marginalization. And by the nature of storytelling, the PDFis gave every storyteller the opportunity to include details that were not task-oriented (as in a scenario \cite[section 3.7.3]{oliver2021obstacles}) but did contribute to an enhanced understanding of the participant/author \cite{baumer2020evaluating}. When Ahmadpour et al. \cite{ahmadpour2019co} asked their participants to critique DFs written by the researchers: ``We found that the participants had responded to some contextual elements of the fiction that were unrelated to any specific concept.'' Whether or not participants express themes of marginalization in their stories, the value of DF to user studies was recognized by Baumer et al. \cite{baumer2020evaluating}, in their paper about the ``epistemological confusion'' that clouds evaluation of a given work's potential contribution.

At this point, you are probably asking one question:

\subsection{What is the contribution?}
Either I failed to state the contribution clearly enough in the previous section, or I have failed to make a persuasive case for the significance of the contribution. It happens all the time.

Sure seems difficult to get across, for some reason. I think I've stated the contribution in words of one syllable, and that is the cue for someone to ask ``So, what's the contribution? I don't get it.''

\section{OPINION AND PRECEDENT VS. THE DEMON REVIEWER 2}
The problems arising from unintended consequences of the anonymous review process, as described in section 4.1, are frustrating, but they are understandable. 

A different kind of problem, when arguing for the value of this methodology to the field of HCI, is that everything I say is a matter of opinion. I will never be able to state it as fact. 
I could, for example, say ``mainstream wearables do not offer a lot of emotional engagement'' and support that with a lot of argument \cite{curtis2014where} \cite{maragiannis2019diversity} \cite{mitchell2014shackles} \cite{liao2020revealing}, but someone else can simply \href{https://www.youtube.com/watch?v=ohDB5gbtaEQ}{gainsay that} (they have) \footnote{\href{https://en.wikipedia.org/wiki/Argument_Clinic}{\textit{Wikipedia:} \textbf{Argument Clinic} https://en.wikipedia.org/wiki/Argument\_Clinic}}. When the person doing the gainsaying is the reviewer, they win. What they say is undoubtedly true from their own point of view; I infer that we have a philosophical difference. What I said is also true from my own point of view - at least, I do not perceive mainstream wearables to be offering the kind of emotional engagement that would make them interesting to me; and I share that perception with a number of experts in both industry and academia, whom I could cite \cite{hunn2015the} \cite{maciocci2013me} \cite{mcgookin2014wearables} \cite{wallace2007emotionally}, but I will not get the chance to, because reviewer says no.

I can say: I used this method of PDFi and I got these results. And someone else can ask: how do you know that the PDFi is what got those results? \footnote{One could reason that the question is really asking for a defense of PDFi in comparison to other methods. I would have taken it that way, too, had its wording not clearly indicated otherwise.} And that question is impossible to answer. I can point to projects that are similar in spirit \cite{spiel2016embodied} and say: if someone did the same experiment without PDFi, it would look something like this. If and when somebody does that, we might have some idea if PDFi is what made the difference. For example, Spiel et al. 2016 \cite{spiel2016embodied} does something similar, albeit with companion technologies rather than wearables, albeit with a cohort of autistic children, but with similar goals and values; just without using PDFi as a method of concept elicitation. They did not have results at the time of publishing that paper, but it is an example of what a comparable study without PDFi would look like.

One paper that answers, as directly as possible, the question of the relative contribution of PDFi is N\"{a}gele et al \cite{nagele2018pdfi}, who acknowledge ``Although effectiveness of a DF is difficult to quantify [35] there is promise that the PDFi method can surface insights that effect and inspire developers of medical device technologies and rightfully give voice to people in vulnerable positions.'' 

But the basic problem I have with that question is that I never did say PDFi was \textit{better} than other methods. I said it was \textit{a} method of doing something that nobody else has ever done, which is \textbf{ask participants what wearables they want, make the wearables, hand them over, await feedback for one year, reflect with participants about a) further development and b) sociotechnical futures.}

\section{SO WHAT?} 
Underlying this work is my assumption that discovering what kinds of wearables are desirable to potential users, is actually important. This seems self-evident to me, but let me attempt to justify my position: As I explained in section 2, if you already know your participants use fitness monitors, it makes sense to study the co-design of ways to style fitness monitors. However, also as described in section 2, most participatory design studies of wearables are based on technology push. Participants might enjoy co-designing your smart glasses \cite{cassidy2015participatory} in the workshop, but does that mean they will want smart glasses when the workshop is over? If a participant comes up with their own idea, they might or might not like the real-world implementation of that idea, but does it not stand to reason that the exploration of the participant's own idea would provide actionable insights that we might not get from the ``push'' approach? 

The question of what kinds of wearables are desirable to potential users \textit{has} been asked \cite{fortmann2015user} \cite{mueller2005hug} \cite{liao2020revealing} - it is not novel in itself. But as I explained in section 4, no-one has tried to answer it in this particular way. 

So: in what scheme of things does the combination of this specific question (of how to get nonexpert participants to come up with concepts for everyday connected wearables) and this specific way of answering it (ask them to tell a story about the wearables they want, make those wearables, reflect after a time in-the-wild) amount to a methodological contribution to HCI, and not just a one-off? Within what epistemic framework is this study important, and not incidental? 

\subsection{I'd like to have an argument, please}
I need to elucidate a theoretical framework for WHY this is important. Why indeed? Why is it important to find out what wearers want? Why is asking them a good way of finding out? Why is PDFi - that is, co-creation through storytelling - a good way of finding out? However clearly I try to state my case, Reviewer 2 can always dismiss the whole elephant by focusing on the elephant's toenail and ignoring the rest. 

So, one more time: \textit{has} anybody done participatory design of wearables like this before? As I have shown in section 4.2: Jones et al. had an open-ended brief, but their sights were fixed on a future horizon. My own work was aimed at taking action in the present in hope of democratizing the future of everyday wearables. But it takes more than that to explain the novelty and contribution of my use of participatory design. 

Has anybody done \textit{PDFi} like this? No, but it turns out not to matter, because Reviewer 2 is an expert in DF and says that this is \textit{not} DF. When Lindley and Coulton \cite{lindley2015back} ``posit[ed] that design fiction is inherently flexible'' I took that to mean that experimentation with the genre would be acceptable. Reviewer 2 takes it to mean that contributions are admitted or excluded on a case-by-case basis, and wields an ad hoc definition of what DF is \textit{not} like a flick knife in a dark alley. 

For example, Reviewer 2 asserts that DF is NOT storytelling. That assertion is trivially refutable through the literature, because the question of whether DF is storytelling has been asked, and the answer from the World Design Fiction Council \footnote{Not a real council.} was that it \textit{can} be \cite{coulton2017design}. 
Of course this too is a matter of opinion, but it is an opinion that is expressed in The Literature. The \textit{fact} of the matter will never be something I or anyone can prove (at least according to Dindler \cite{dindler2010fictional}), but I \textit{can} prove that there is a spot in The Literature of DF that says it \textit{can be} storytelling \cite{coulton2017design}, even if storytelling is not preferred. There it is, right there in The Literature. I don't make the rules. 

I could \textit{try} making the rules. I could try making the case that DF was not storytelling before, but NOW it is. I \textit{am} trying to make the rule \cite[section 3.7.3]{oliver2021obstacles} that DF was not for design in the present \cite{lindley2015a} until recently \cite{jensen2017ethical}, but NOW it can be.

I would be no more wrong or right either way, but sadly my rightness or wrongness is irrelevant because I do not have that much influence. I do not have the influence because I do not have enough publications, and I cannot publish because I cannot fit my work into any existing theoretical framework. 

Or: I \textit{can} fit my work into an existing theoretical framework, and I can back up that assertion with citations, but Reviewer 2 has left the building and I am only talking to myself. I turn to one subdiscipline after another, like an orphaned duckling in search of a mother.

I could say: look, see the wearable design artefacts resulting from this research. But they are not technologically very innovative, nor are they well made (they are technology probes \cite{hutchinson2003technology}, not finished products). To the casual observer, the work looks like it could be based in the visual arts. But conferences in the domain of visual arts, which ask for visual and graphic contributions, would not care for these because these are illustrations of a process, they are not a finished product. For the same reasons - they are not well executed and do not contribute to the technospace of wearable device development - no wearables venue is going to look at them either. 

\section{THE UGLY DUCKLING IS STILL WAITING}
But the point is these are what wearers of wearables have indicated that they want. And they show a path for wearables designers to overcome various obstacles to wearable computing. 

But Reviewer 2 says case studies are not a good method for this kind of study. Case studies are good enough for others in this domain, though \cite{wallace2007emotionally}? Whatever the reasoning behind the rejection of case studies, that particular door to wearables as a subdiscipline is closed to me.

As far as HCI is concerned, I am simply doing old-hat boring participatory design, combined with some erroneous idea of DF that is not DF, and therefore I am not making the contribution I claim... whatever that contribution even is, they can't quite tell from what I have written.

What about the argument that DF is not for taking action in the present? I left that objection dangling in section 6.1. I could point to Jensen and Vistisen \cite{jensen2017ethical}, who argue for DF that turns to action in the present as a way of incorporating ethics in the design process. Reviewer 2 says it's not, and I may be looking into the abyss of yet another philosophical difference. Baumer et al. \cite{baumer2020evaluating} have argued for greater open-mindedness and more willingness to acknowledge the validity of different schools of thought about ways to apply DF (``what is the value that members of a community wish to take from design fiction?''), but so far I am singing Kum Ba Yah \footnote{\href{https://en.wikipedia.org/wiki/Kumbaya}{\textit{Wikipedia:} \textbf{Kumbaya} https://en.wikipedia.org/wiki/Kumbaya}} on my own. 

I have half-joked \cite[section 3.7.3]{oliver2021obstacles} that if logical defenses of my own application of DF continue to cut no ice, I will call it something else and move on. Perhaps I was not joking. In 2017, at which point I had been working with PDFi for a year or two, Elsden et al. \cite{elsden2017on} wrote about their own multiple years of work in ``speculative enactments''. The paper provides a baroque framework with many detailed criteria for what constitutes a speculative enactment; clearly my own work does not match their description de jure, though de facto it has much in common, in particular the advocacy for consequential participation. Their main point is that the time has come to start DOING things with participants, not limiting their involvement to talking or thinking. I can point to them and say: look, it is not just me. More and more people are arguing for action and not just talk.

Cue Frauenberger \cite{frauenberger2020entanglement} on ``Entanglement HCI'', expressing considerable angst about Actor-Network Theory, Object-Oriented Ontology and Postphenomenology. Here is my opportunity to show that the main design artefact of my own research - the Gallery Necklace \cite{oliver2019design} - is in a totally different relation to the wearer than Ihde \cite{ihde1990technology} ever even noticed. Ryan’s 2014 book \cite{ryan2014garments} about the history of wearables remarked on the postphenomenologist’s inattention to wearables, with a sidelong glance at Barthes into the bargain. Ihde's position is that clothing is in the background of an embodiment relation with the wearer, in which the human relates to the world \textit{through} technology without paying attention to the technology itself \cite[section 3.6.2]{oliver2021obstacles}. 

Ihde's concept of an \textit{alterity} relation is one in which the human's attention \textit{is} on the technology itself. A toy would be in an alterity relation with the human. And the Gallery Necklace is too \cite{oliver2019design}. The Gallery Necklace is an eInk screen, wearable in the setting of a statement necklace, with an image that changes every three minutes when powered on. The purpose of the Gallery Necklace is to display the wearer's own artwork as a conversation piece and icebreaker for in-person social interaction. The Gallery Necklace is an item of technology that puts the human's attention onto itself. Yes, the ultimate goal is to draw attention to the human wearing it, but that does not really mean it is in the same kind of embodiment relation we have with a phone. You talk to the world through the phone and you connect with the wearer through the Gallery Necklace, but one thing is not like the other because you must pay attention to the Gallery Necklace itself as an essential part of the process. Mainstream wearables do not offer alterity relations: they are styled in such a way as to be worn in a background embodiment relation. They also provide hermeneutic display of the wearer's own biodata. 

The Gallery necklace subverts all the expectations that mainstream commercial wearables have built up in our minds. The Gallery Necklace \textit{is} in an alterity relation and \textit{does} draw attention to itself, and the Gallery Necklace is what participants have indicated that they want. Instead of biodata, it displays artwork - in a way, you could say it displays what is in the wearer's mind, but it is not hermeneutic \cite[section 3.4.1]{oliver2021obstacles}. 

The version we have is not a finished product, so is not suitable for visual arts-oriented or wearable technology venues - but it \textit{is} what participants wanted. They want a more polished implementation of the Gallery Necklace \cite[section 5.2]{oliver2021obstacles}, and some of them want it as a Badge instead of a necklace, but the point is that one participant came up with the concept for the Gallery device \cite[Appendix E.7]{oliver2021obstacles}, the whole group asked for a Gallery device \cite[section 4.2]{oliver2021obstacles}, they got it, and they want it. That is, they want the Gallery Necklace done better - as opposed to some completely different device. This outcome supports the potential of PDFi as a source of inspiration for the user-centred design of everyday connected wearables.

It is curious that Ihde would skip the possibility of alterity relations where clothing is concerned, and thereby leave out so much about the ways humans have related to clothing. This \href{https://youtu.be/YOmJ54WpUVg}{Gravettian man}, buried in clothing embellished with thousands of tiny shells and teeth - is it enough to describe this as an embodiment relation? There was someone who wanted to look stylish and rich and powerful, and he did. Perhaps the embellishments were part of a costume of office, in which case you could say that you look through the embellished clothing at the social role of the wearer. But that would only mean I had chosen the wrong example and should pick someone else. Wearable items are as much about drawing attention to themselves, for themselves, as they are for embodiment or (in the case of smartwatches) hermeneutics. Engineers forget \cite[section 3.6.3]{oliver2021obstacles} what neolithic icemen remembered. 

At this point, someone can always assert that I am wrong about this, and say the Gallery Necklace is not in an alterity relation with the wearer, and/or if it is that is not novel, and so on. They might or might not have some valid arguments about that particular point. 

But, again, Frauenberger \cite{frauenberger2020entanglement} is adding his voice to those who say we need action, not just speculation \cite{jensen2017ethical} \cite{elsden2017on}. According to Frauenberger, we need a new paradigm, we need speculative enactments, we need agonistic design to get conflicts out in the open \footnote{And may I suggest that a workshop of participatory storytelling, that aims to find a way of honouring every individual perspective to the extent possible with the available time and resources, would be a perfect opportunity for agonistic design.}, we need to deemphasize user-centered design. According to Frauenberger, it is time to start designing for meaningful relations with technology. He calls this Meaningful Design.

If nothing else, Frauenberger's proposal of Entanglement HCI offers me some new keywords so that Reviewer 2 cannot dismiss my work on the grounds that it is not DF - or they can, but not for that reason. One down, 999 to go. 

If I decide to call my PDFis something else in order to keep going, I will have to choose from the existing jargon, perhaps resulting in nothing better than rejection from another subdiscipline. Unlike Frauenberger, who has standing in the community, I cannot tell the HCI community that what it really needs is a new paradigm with new keywords. But now that he has said what I was thinking, at the late date of 2020, I can finally show that my own work fits into this new paradigm and prove that I have the receipts to \textit{show} that which Frauenberger is \textit{telling}.

I just cannot state it in so many words. As explained in section 1, I was motivated initially by my own dissatisfaction with the state of mainstream wearables, and of course Reviewer 2 has rejected this motivation as insufficient. 

\section{ARE YOU STILL HERE?!?}
Why am I still insisting on the importance of this? If six people tell you you're dead, lie down.

Except that it's not only about me.

In section 4.1, I described my intention to write an in-depth thematic analysis of the PDFis. I believe the PDFis, in their full-text versions \cite[Appendices A--G]{oliver2021obstacles}, are not simply data available for extractive reading \cite{liboiron2020exchanging} at best, interpretation through an academic authority figure (or, until one comes along, me) at worst. I believe they are design artefacts in themselves, and worth reading in and of themselves.

I believe that the PDFis, being obviously deeply felt creative expressions by my volunteers, and including some expressing the pain of ongoing marginalization, are worthy of the direct attention of the research community. 

It pains me when they are met with condescension, as they have been, more than once: oh, what nice little stories. But you see dear, this is an academic research venue, we do science here, you see. But you know these little stories are really ever so nice! Perhaps find a venue with lower standards?

Perhaps it was unrealistic to expect better - after all, that is what being marginalized means.

But perhaps the research community is a little too comfortable setting limits on what their participants can say and what they are willing to hear. Which, now that I think of it, was my point in the first place.

\section{CONCLUSION}
This is not the end.

\section*{ACKNOWLEDGEMENTS}
Thanks to The Alan Turing Institute for funding and supporting this doctoral study; and to Jon Crowcroft and Richard Mortier for egging me on to submit this; and to Anil Madhavapeddy and Irina Shklovski for being the Anti-Reviewer2.

\bibliographystyle{abbrv}
\bibliography{effort}

\balance

\end{document}